\documentclass[a4paper, onecolumn]{article}

\usepackage[pages=all, color=black, position={current page.south}, placement=bottom, scale=1, opacity=1, vshift=5mm]{background}
\SetBgContents{
}      

\usepackage[margin=1in]{geometry} 

\usepackage{amsmath}
\usepackage{amsthm}
\usepackage{amssymb}

\usepackage[utf8]{inputenc}
\usepackage{hyperref}
\hypersetup{
	unicode,
	pdfauthor={Author One, Author Two, Author Three},
	pdftitle={A simple article template},
	pdfsubject={A simple article template},
	pdfkeywords={article, template, simple},
	pdfproducer={LaTeX},
	pdfcreator={pdflatex}
}

\usepackage[sort&compress,numbers,square]{natbib}

\theoremstyle{plain}

\theoremstyle{definition}

\usepackage{graphicx, color}
\graphicspath{{fig/}}

\usepackage{algorithm, algpseudocode} 
\usepackage{mathrsfs} 

\usepackage{lipsum}

\title{Self-Organized Freeform Waveguiding}
\author{Fadhila Chehami$^{1*}$ \and Cyril Decroze$^1$ \and David R. Smith$^2$ \and Thomas Fromentèze$^1$ }

\date{
	$^1$ \texttt{University of Limoges, XLIM, UMR 7252, F-87000 Limoges, France}\\%
	$^2$ \texttt{Center for Metamaterials and Integrated Plasmonics, Department of Electrical and Computer Engineering, Duke University
	, Durham, North Carolina 27708, USA}\\
    $^*$ fadhila.chehami@unilim.fr
    }

\begin{document}
	\maketitle
	
		
\noindent\textbf{Abstract:}
	
Nature offers remarkable examples of complex photonic architectures such as those responsible for the iridescent colors of butterfly wings that emerge spontaneously during growth, well before any centralized control takes place. Arising from local rules, these structures exhibit advanced optical functionalities, such as photonic band gaps, without relying on in-situ optimization or top-down design. Inspired by biological morphogenesis, we introduce an optimization-free approach for the automated generation of self-organized freeform waveguides that adapt to complex propagation paths. Our method relies on local reaction-diffusion dynamics to produce robust, spatially distributed structures. In contrast to conventional waveguides based on periodic media, which impose strong geometric constraints and require extensive fine-tuning, the proposed structures support nontrivial geometries while maintaining photonic band gap behavior. We experimentally demonstrate that these self-organized waveguides achieve superior transmission efficiency along complex paths. This optimization-free strategy enables the automated design of advanced electromagnetic components with intrinsic adaptability and resilience.\\
		
Optimization-free materials\\

\section{Introduction}
Efficiently guiding electromagnetic waves through complex, arbitrary geometries is a critical challenge in modern photonics, particularly for integrated photonic circuits and advanced optical communication systems~\cite{cao2022shaping, li2021directional}. Conventional optical waveguides, typically optimized for straight or mildly curved paths, exhibit significant performance deterioration when routing signals through more intricate trajectories~\cite{ishizaki2013realization, song2020low}. To address this challenge, photonic crystals, periodic materials capable of supporting electromagnetic band gaps (EBGs) where wave propagation is prohibited, have emerged as a powerful alternative, offering tight wave confinement and precise propagation control~\cite{yablonovitch1987inhibited, joannopoulos1997photonic}. However, achieving robust guidance through complex paths within these periodic structures often remains elusive due to their inherent sensitivity to geometric deviations, symmetry breaking, and fabrication imperfections~\cite{zhou2014ultra, yuan2016design}.

Numerous optimization-based approaches have been proposed to mitigate losses arising from structural irregularities in photonic crystals. These techniques range from parametric tuning of nearby elements~\cite{askari2010systematic, jiang2013optimization}, addition of strategically placed auxiliary scatterers~\cite{hu2011improved}, resonant cavity formation for impedance matching~\cite{mekis1996high, boscolo2002numerical}, to sophisticated topology optimization methods reshaping lattice elements into intricate geometries~\cite{tee2014high, chen2012design}. While successful in specific scenarios, these strategies inherently rely on intensive computational processes, exhaustive case-by-case tuning, or impose substantial fabrication complexity, ultimately limiting their broader applicability and scalability.

\begin{figure}[H]
	\centering
	\includegraphics[width=0.9\textwidth]{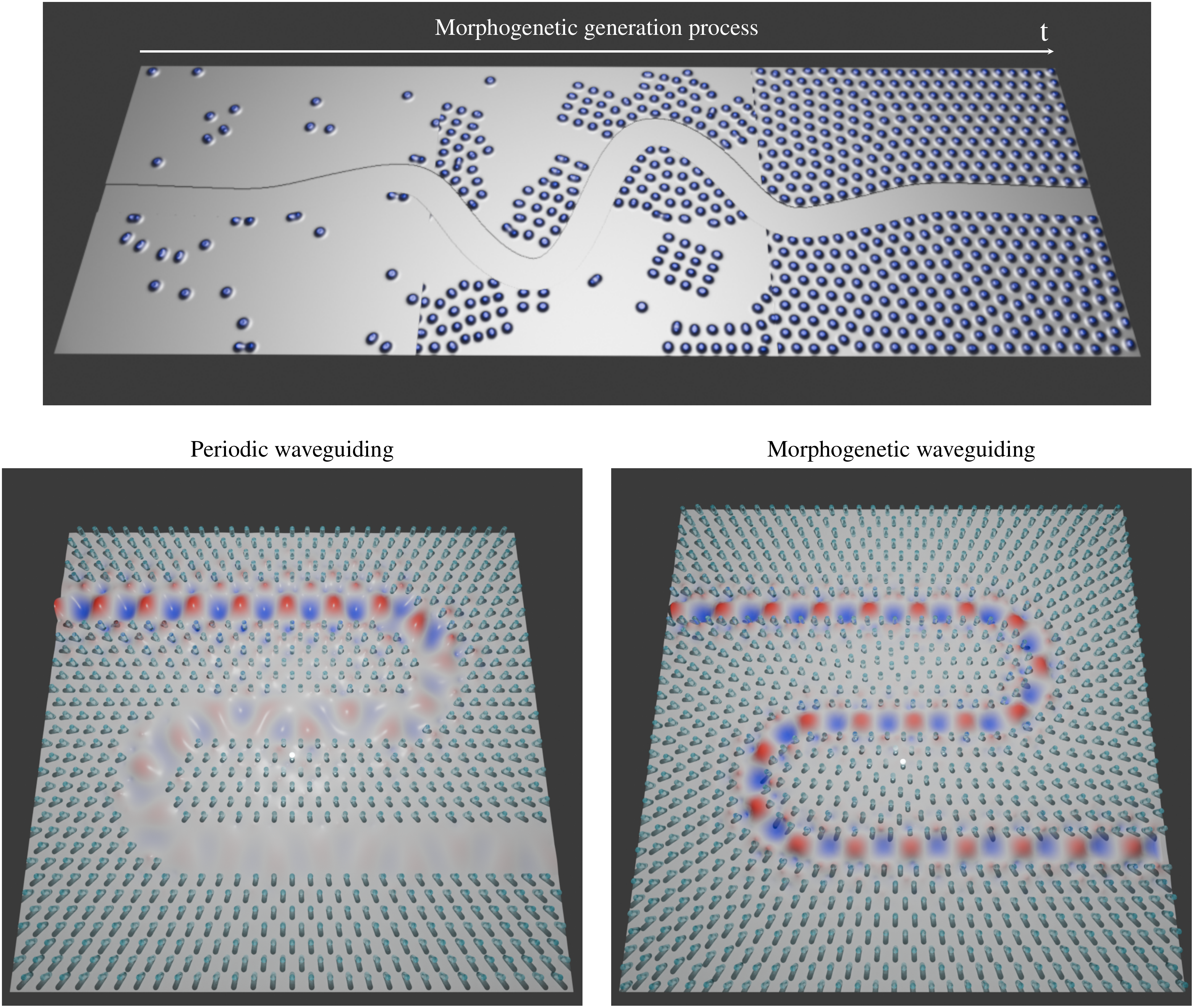}
	\caption{Freeform waveguiding enabled by morphogenetic generation. Top panel: Temporal evolution of the morphogenetic generation process. Circular patterns progressively self-organize into a disordered hyperuniform medium embedding a predefined freeform guiding path, without the need for global optimization or centralized control. Bottom panel: Simulated electric field distributions in two bent waveguiding structures. The left panel shows a conventional triangular photonic crystal where the path is carved by element removal, strong scattering and reflections are observed at the bends. The right panel illustrates the morphogenetic counterpart, in which the wave smoothly follows the complex path with minimal loss, highlighting the adaptability of the self-organized structure.}
	\label{fig:Generation_Guiding}
\end{figure}

Here, inspired by biological pattern formation and morphogenesis theory introduced by Alan Turing~\cite{turing1952chemical}, we propose a fundamentally different, optimization-free approach for designing broadband freeform waveguides based on self-organized hyperuniform disordered structures (Fig.~\ref{fig:Generation_Guiding}). Hyperuniform media inherently suppress density fluctuations at large scales, offering isotropic band gaps without requiring structural periodicity~\cite{torquato2015ensemble}. Through local reaction-diffusion dynamics guided by simple interaction rules, our morphogenetic method spontaneously generates robust and adaptable patterns, naturally conforming to arbitrary guiding paths with exceptional fidelity and minimal losses. This strategy represents a significant paradigm shift from computationally demanding, centralized optimization towards decentralized, self-organized photonic structure generation potentially transforming waveguide design with unprecedented flexibility and resilience.

Although Turing’s morphogenesis theory was introduced as early as 1952~\cite{turing1952chemical}, its broader scientific significance only gained recognition several decades later, driven by extensive experimental validations and refined computational models~\cite{kondo2010reaction}. Initial explorations beyond biochemical contexts, notably by Doursat \textit{et al.}, proposed leveraging decentralized self-organization as a general framework for designing complex, functional structures across multiple scientific and engineering disciplines~\cite{doursat2013review}. 

Building on this principle of decentralized synthesis, our group has investigated the application of morphogenetic processes to metamaterial-based electromagnetic systems. We notably introduced the concept of morphogenetic metasurfaces, where self-organization under geometric constraints gives rise to complex subwavelength architectures with tailored tensorial properties~\cite{fromenteze2023morphogenetic}. This approach circumvents the need for strict periodicity or symmetry, relying instead on the intrinsic degrees of freedom inherent to pattern self-structuring to enhance effective electromagnetic responses. We have also explored the development of generative models for secure identification, encoding structural singularities directly into antenna radiation patterns~\cite{amorim2023morphogenetic}. In parallel, we showed that the same dynamics can produce disordered yet hyperuniform media supporting isotropic photonic band gaps~\cite{chehami2023Eucap, chehami2023ACS}. This ability to link local morphogenetic structuring with robust spectral behavior forms the basis of the present work, where bandgap formation and waveguiding emerge jointly through a fully self-organized, geometry-adaptive process.

The present work builds upon the seminal contribution of Man \textit{et al.}~\cite{man2013isotropic}, who demonstrated the feasibility of waveguiding through complex geometries using hyperuniform disordered materials. Their approach, based on gradient-descent optimization over a high-dimensional design space, effectively tailored local disorder to enable wave propagation. However, in their experimental implementation, the guiding structures did not fully conform to the imposed curved paths, particularly near the boundaries, where irregularities and mismatches remained visible. 

In contrast, the morphogenetic process introduced in our work offers a fundamentally different strategy. By leveraging local reaction-diffusion dynamics, the generated structures exhibit a spontaneous and coherent adaptation to imposed boundary constraints. The resulting patterns remain smooth and uniform, even along strongly curved paths. The approach is fully decentralized, requires no global optimization, and remains intrinsically scalable making it particularly suitable for the automated design of large-scale, freeform waveguides.

In the remainder of this work, we first detail the generative process and benchmark its performance in the Results section, then discuss its broader implications and potential extensions in the Discussion, and finally provide implementation specifics in the Methods, which offer a concise synthesis of the procedures fully described in the Supplementary Materials.

\section{Results}
    
\subsection*{Morphogenetic generative technique}

Turing’s morphogenesis theory attributes the spontaneous emergence of patterns in nature to the interplay between reaction and diffusion processes~\cite{turing1952chemical}. In this work, we implement a generative model based on the Gray–Scott reaction–diffusion dynamics, which provide a robust framework for the autonomous formation of spatially structured patterns~\cite{gray1983autocatalytic}. The model is governed by a set of coupled partial differential equations (Eq.~\ref{eq:gray-scott}) that describe the nonlinear interaction between two chemical species, $U$ and $V$, leading to the emergence of diverse morphologies~\cite{kondo2010reaction}. Details of the numerical implementation are provided in Section 1 of the Supplementary Materials.

\begin{equation}
	\begin{aligned}
		\frac{\partial \mathbf{U}}{\partial t} &= d_U \nabla^2 \mathbf{U} - \mathbf{U} \mathbf{V}^2 + f(1 - \mathbf{U}) \\
		\frac{\partial \mathbf{V}}{\partial t} &= d_V \nabla^2 \mathbf{V} + \mathbf{U} \mathbf{V}^2 - (f + k)\mathbf{V}
	\end{aligned}
	\label{eq:gray-scott}
\end{equation}

\begin{figure}[H]
	\centering
	\includegraphics[width=0.8\textwidth]{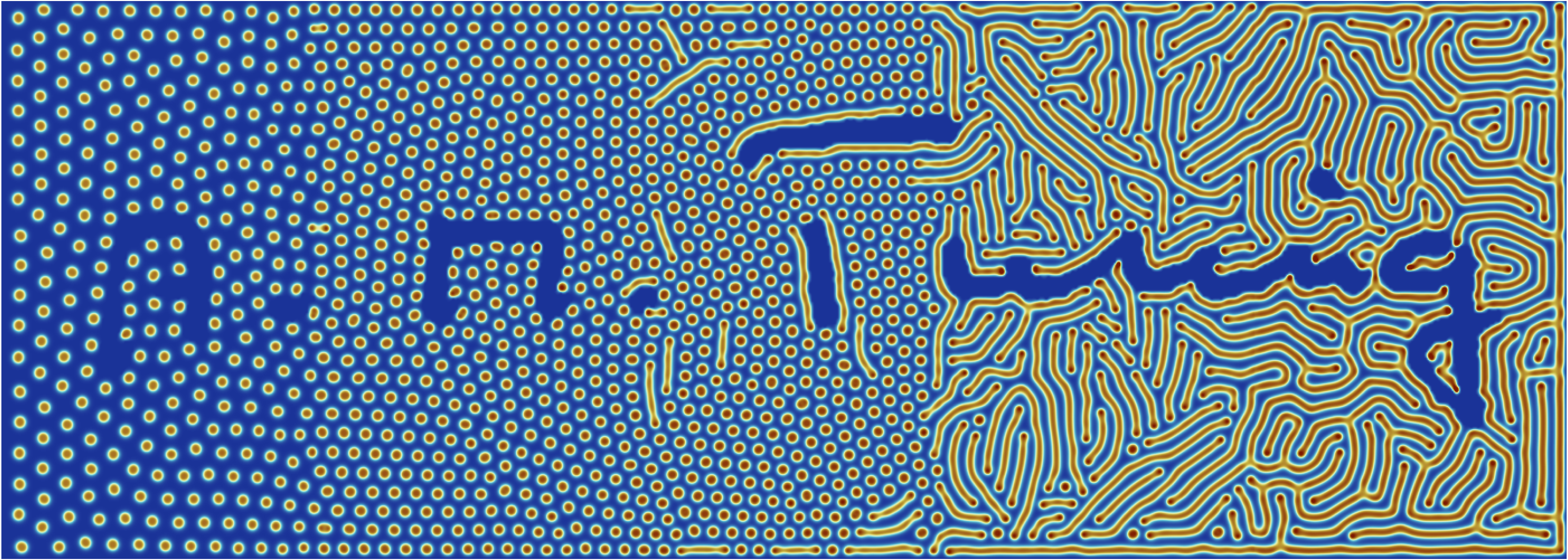}
	\caption{Pattern generation under geometric constraints using reaction–diffusion dynamics inspired by Turing’s morphogenesis theory. The imposed boundary condition corresponds to Alan Turing’s handwritten signature, illustrating the spontaneous adaptation of morphogenetic patterns to complex shapes.} 
	\label{fig:Morpho_theory}
	\end{figure}

The reaction term corresponds to the transformation of a $U$ particle into $V$ in the presence of two $V$ particles, while the diffusion of each species occurs independently, governed by the diffusion coefficients $d_U$ and $d_V$ through the Laplacian operator $\nabla^2$. A dynamic equilibrium is maintained through two additional parameters: the feed rate $f$, which drives the spontaneous production of $U$, and the kill rate $k$, which accounts for the decay of $V$. Depending on the choice of parameters and initial conditions, the Gray–Scott model can generate a wide variety of patterns, which have been systematically classified by Pearson~\cite{pearson1993complex} (see also Fig.~\ref{fig:Morpho_theory}). An extended overview of the diversity of emergent Turing patterns is available in Section 1 of the Supplementary Materials. Interactive simulations of such dynamics can also be explored online via the VisualPDE platform~\cite{walker2023visualpde}, which provides real-time visualization of reaction–diffusion systems, including the Gray–Scott model. 

In the present work, we focus on the generation of circular spot patterns, obtained for parameters $f = 0.036$, $k = 0.065$, $d_U = 1$, and $d_V = d_U/2$. These configurations are used to define spatial distributions of dielectric cylinders exhibiting isotropic electromagnetic band gaps (see Supplementary Section 2). The resulting patterns can be harnessed to define spatial layouts of dielectric inclusions, bridging reaction–diffusion dynamics and photonic structure design.

\subsection*{Freeform waveguide design}

To ensure a fair comparison between morphogenetic and periodic architectures, both media were first dimensioned in their full, defect-free configurations. The design parameters such as lattice constant and cylinder radius were selected based on photonic band diagram simulations, with the objective of obtaining a complete band gap within the target frequency range. This guarantees that the resulting materials do not support propagating modes in the absence of a guiding path, thereby enabling efficient confinement once a waveguide is introduced. 

\begin{figure}[H]
	\centering
	\includegraphics[width=0.82\textwidth]{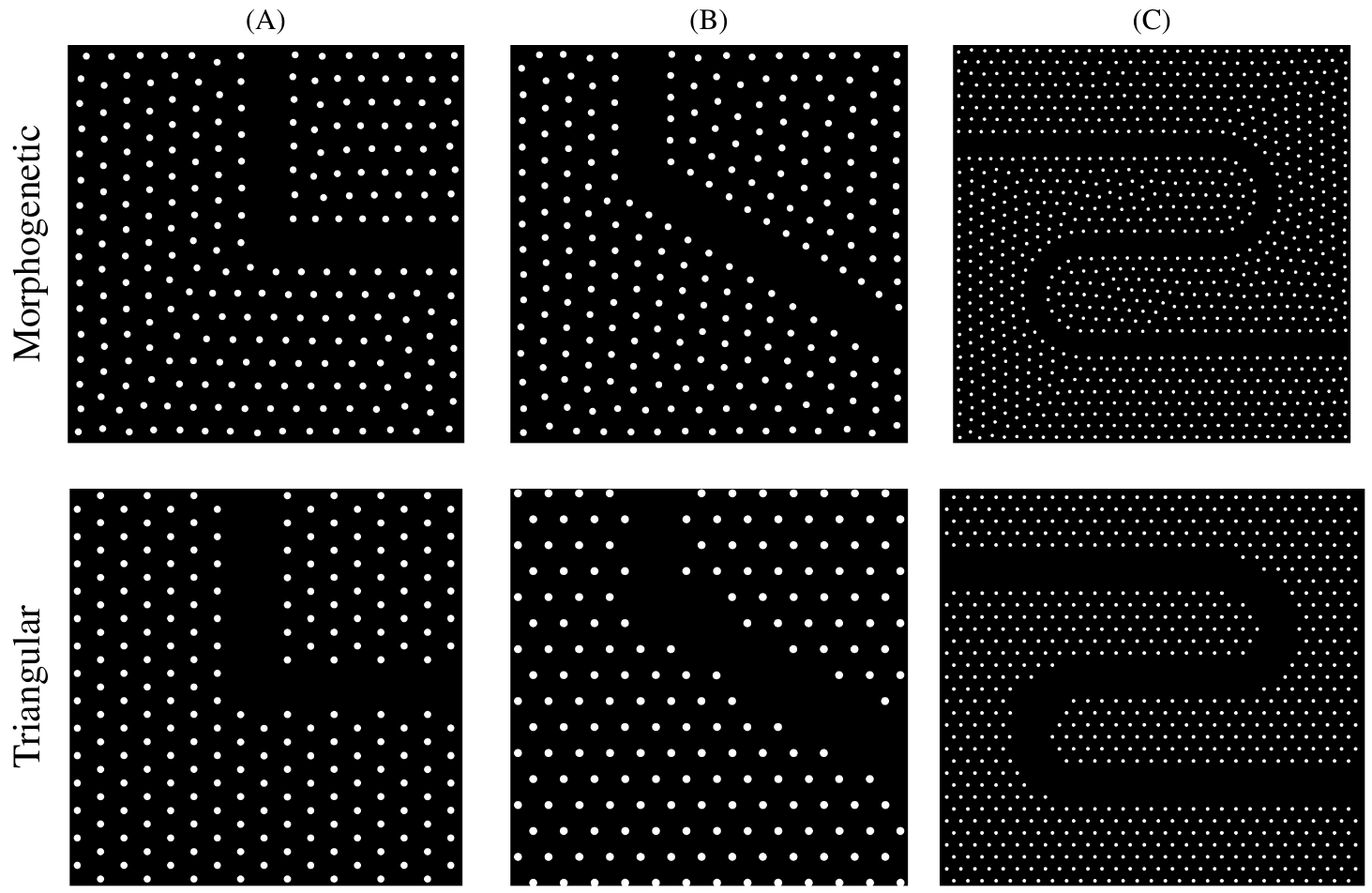}
	\caption{Top views of the freeform waveguides investigated in this study: (A) 90° bend, (B) 120° bend, and (C) S-shaped waveguide. The structures consist of cylindrical alumina patterns ($\epsilon_r = 9.8$, $\tan(\delta) = 2 \times 10^{-3}$) with a radius of $0.85$\,mm embedded in air. The samples have been generated with $f=0.036$, $k=0.065$, $d_U=1$, $d_V=d_U/2$ and $\rho = 0.9$ within a $300 \times 300$ pixel domain for the 90° and 120° bends and a $600 \times 600$ pixel domain in the case of the S-shaped waveguide. The upper row corresponds to morphogenetic disordered distributions, while the lower row shows periodic triangular crystal arrangements.}
	\label{fig:Echantillons_90_120_S_Morpho_Triang}
\end{figure} 

As illustrated in Fig.~\ref{fig:Echantillons_90_120_S_Morpho_Triang}, three representative freeform waveguide geometries were considered namely, a 90° bend, a 120° bend and an S-shaped curve. These configurations are commonly used as benchmarks in the literature and have motivated the development of various optimization techniques. The resulting structures reveal that, in the morphogenetic case, the patterns self-organize with high precision, closely following the imposed guiding path (An animation illustrating the morphogenetic generation of the S-shaped waveguide is available in Supplementary Movie 1). In contrast, the periodic designs exhibit structural irregularities and a broader guide width, stemming from the fixed lattice spacing and the discrete removal of elements along the desired path.

\begin{figure}[H]
	\centering
	\includegraphics[width=0.78\textwidth]{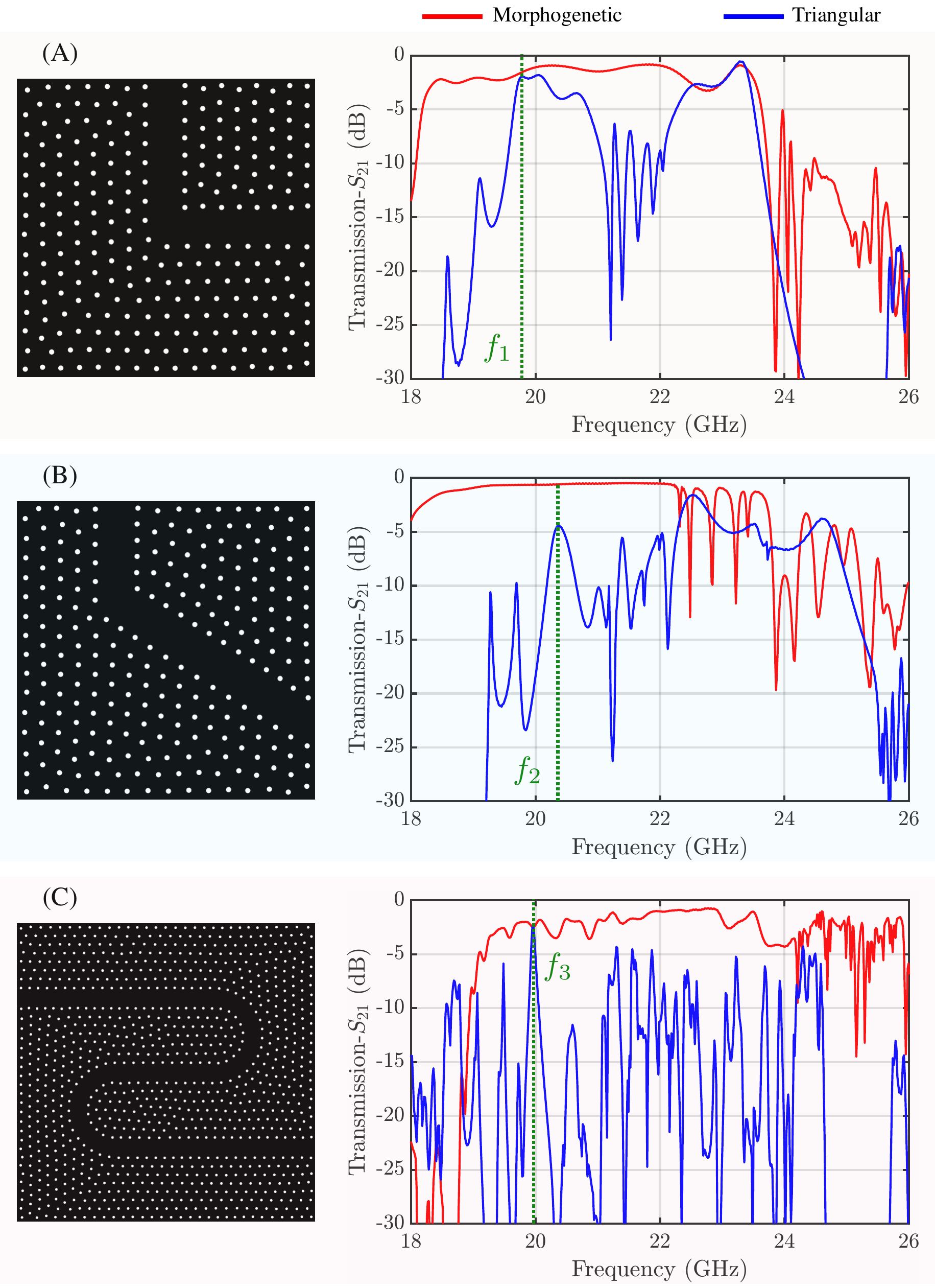}
	\caption{Simulated S$_{21}$ transmission through the freeform waveguides shown in Fig.~\ref{fig:Echantillons_90_120_S_Morpho_Triang}, comparing morphogenetic (red) and triangular (blue) configurations. Insets display the corresponding geometries. Frequencies highlighted in green indicate the peak transmission points: $f_1 = 19.8$\,GHz, $f_2 = 20.4$\,GHz and $f_3 = 19.9$\,GHz.}
	\label{fig:ParamS21_90_120_S_simu}
\end{figure} 

These differences arise from the design methodology. In the morphogenetic case, the guiding path is imposed as a forbidden region during the pattern generation, allowing the structure to adapt naturally to the target geometry. For the triangular lattice, in contrast, all elements intersecting the desired path even partially are removed post hoc (see Supplementary Section 3).

Figure~\ref{fig:ParamS21_90_120_S_simu} presents the simulated S$_{21}$ transmission parameters for each waveguide configuration. The simulations were performed using the method described in Section 5 of the Supplementary Materials. The results clearly indicate that the morphogenetic waveguides achieve substantially higher transmission levels than their triangular counterparts across all geometries. Notably, this enhanced performance is maintained over relatively broad frequency bands.

\begin{figure}[H]
	\centering
	\includegraphics[width=0.9\textwidth]{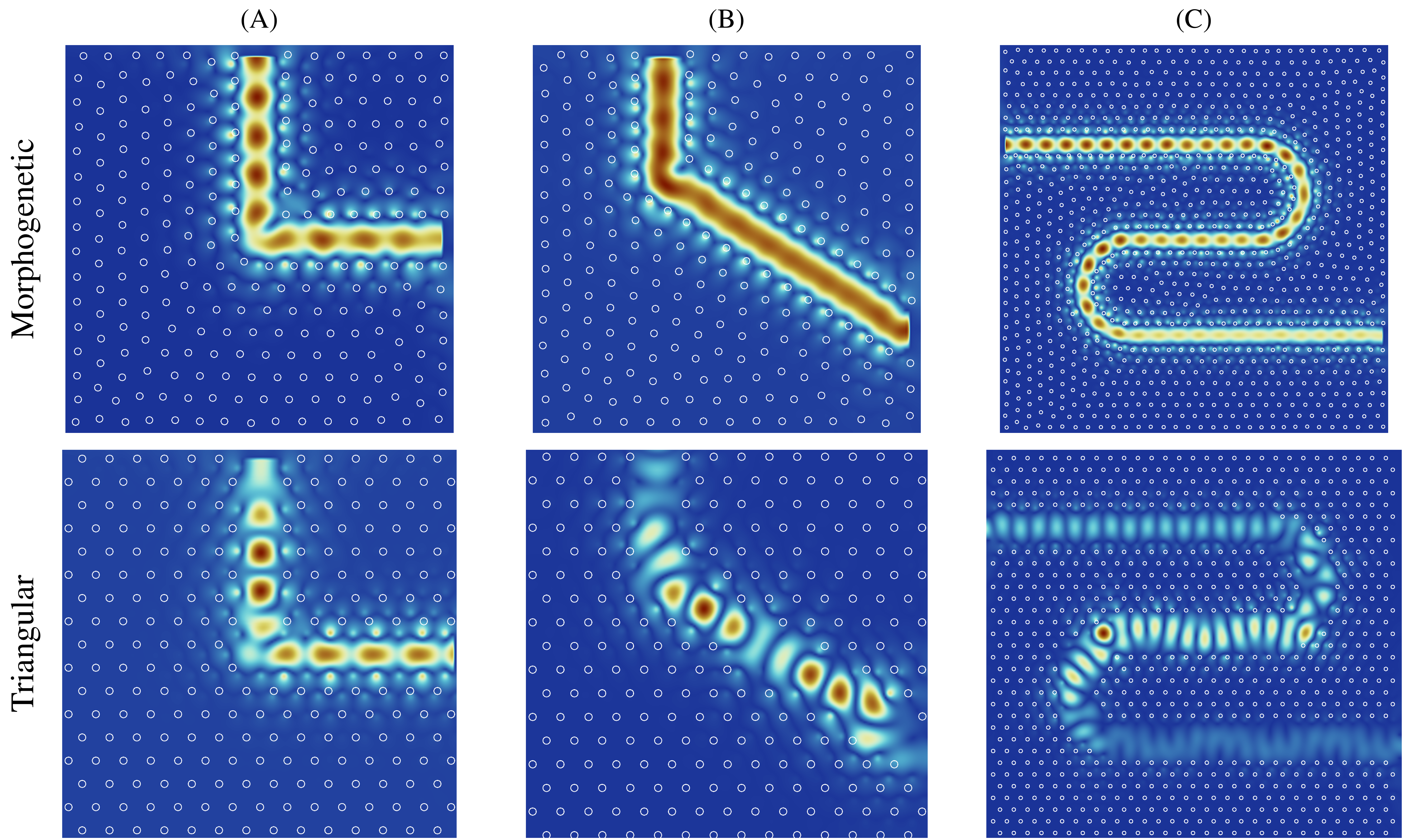}
	\caption{Spatial electric field maps at the frequencies indicated in Fig.~\ref{fig:ParamS21_90_120_S_simu} for the three configurations: (A) 90° bend, (B) 120° bend and (C) S-shaped waveguide. The upper row shows morphogenetic structures while the lower row corresponds to triangular lattices. These fields are respectively extracted at: (A) $f_1 = 19.8$\,GHz, (B) $f_2 = 20.4$\,GHz and (C) $f_3 = 19.9$\,GHz, where the design pairs show joint maximum transmissions.}
	\label{fig:Carto_90_120_S}
\end{figure}

In addition to S-parameter analysis, Fig.~\ref{fig:Carto_90_120_S} displays the corresponding electric field distributions. For consistency, the fields are plotted at the frequencies highlighted in green in Fig.~\ref{fig:ParamS21_90_120_S_simu}, which represent the points of maximum transmission for both the morphogenetic and periodic structures.

These field maps further confirm the superior guiding behavior of the morphogenetic designs: the electromagnetic wave closely follows the intended path with minimal distortion or loss, even through sharp bends. In contrast, the triangular crystal structures exhibit pronounced reflections at curved sections, highlighting their sensitivity to geometric discontinuities (animations displaying electric field propagation through the S-shaped waveguides are available in Supplementary Movie 2 and Supplementary Movie 3, corresponding to the morphogenetic and periodic case, respectively). Such limitations often require complex optimization strategies to mitigate scattering and adapt to fabrication constraints challenges that are inherently avoided with the self-organized approach.

\begin{figure}[h!]
	\centering
	\includegraphics[width=0.9\textwidth]{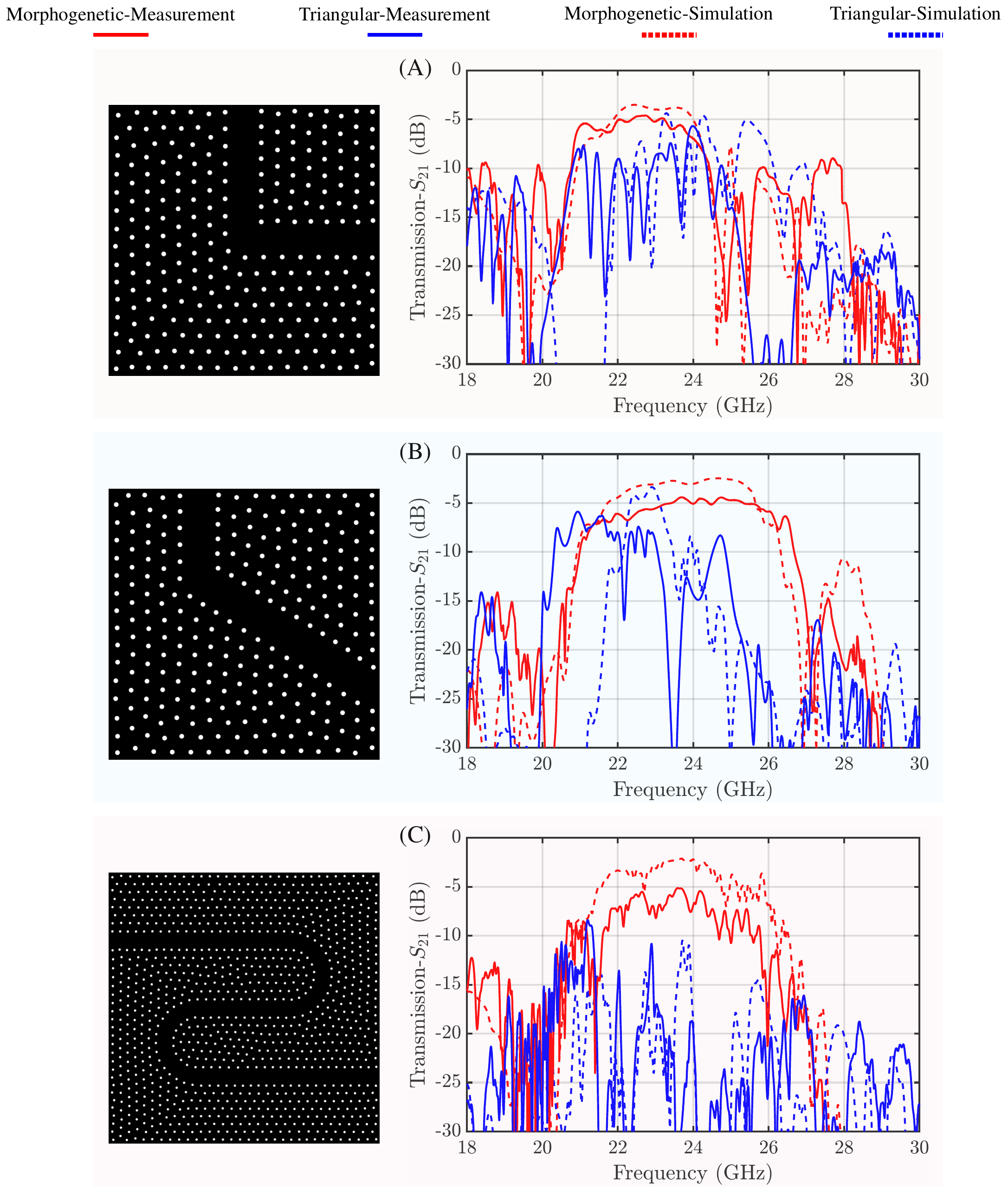}
	\caption{Measured (solid line) and simulated (dashed line) S$_{21}$ transmission parameters corresponding to the proposed freeform waveguides illustrated in Fig.~\ref{fig:Echantillons_90_120_S_Morpho_Triang}. The insets recall the shape of the waveguide considered in each case. The morphogenetic case is plotted in red curves and the triangular case in blue curves. The simulations were performed considering the parameters obtained through reverse engineering namely a \textit{Rohacell} with a relative permittivity of $\epsilon_r$ = 1.07 and an air gap of $0.1$\,mm between the cylinders and the surrounding metallic walls. The measured samples, illustrated in Supplementary Section 4, have overall dimensions of $100 \times 100 \times 5.33$\,mm$^3$ in the case of the the 90° and 120° bends and $200 \times 200 \times 5.33$\,mm$^3$ for the S-shaped waveguide.}
	\label{fig:ParamS21_90_120_S_mesu}
\end{figure}

To assess the practical robustness of our method, we fabricated physical prototypes of the three waveguide configurations and conducted experimental S-parameter measurements. Each structure consists of alumina rods embedded in a low-permittivity \textit{Rohacell} foam matrix, providing mechanical support and well-defined boundary conditions for quasi-two-dimensional wave propagation. The measured S$_{21}$ transmission parameters are shown in Fig.~\ref{fig:ParamS21_90_120_S_mesu}. 

While the spectra confirmed the presence of photonic band gap effects and demonstrated functional waveguiding, the performance of each pair of designs morphogenetic and periodic did not fully match the initial simulation results. In particular, shifts in the band gap frequency and deviations in peak transmission levels were observed. To explain these discrepancies, we carried out a systematic series of simulations based on plausible fabrication tolerances. This led to the identification of a consistent set of corrected parameters that reproduced the experimental trends. Specifically, the best agreement was obtained for a revised permittivity of the \textit{Rohacell} matrix ($\epsilon_r = 1.07$) and the presence of a small air gap of $0.1$\,mm between the alumina rods and the surrounding metallic walls. These corrections, described in the Methods Section and detailed in Supplementary Section 6, enabled the generation of adjusted simulation models that closely matched the measured responses, including the observed frequency shifts and attenuation levels.

The corresponding S$_{11}$ reflection parameter curves, both measured and simulated, are provided in Section 7 of the Supplementary Materials. Despite minor discrepancies and generally low absolute transmission levels primarily due to fabrication tolerances and material variability, the results in Fig.~\ref{fig:ParamS21_90_120_S_mesu} clearly demonstrate the validity of the proposed approach. The morphogenetic waveguides consistently outperform their periodic counterparts, reinforcing the robustness and adaptability of the method under realistic fabrication and measurement conditions.

These results confirm the ability of morphogenetically generated structures to guide waves through complex geometries with high efficiency and minimal optimization. This self-organized approach opens new directions for robust, scalable, and frequency-agnostic photonic device design.

\section{Discussion}

This work introduces the first optimization-free strategy for the design of efficient freeform waveguides capable of operating across broad frequency ranges. Inspired by the principles of biological morphogenesis, the proposed method relies solely on local interactions to generate spatially structured, self-organized media with tailored electromagnetic properties. By circumventing conventional optimization loops, it enables the formation of hyperuniform disordered distributions that naturally adapt to complex boundary constraints. While the absolute band gap width may be narrower than in finely tuned periodic systems, the resulting structures exhibit isotropic band gaps that are significantly more tolerant to geometric distortions and symmetry-breaking defects offering enhanced robustness and greater flexibility for the design of nontrivial guiding geometries.

\begin{figure}[h!]
	\centering
	\includegraphics[width=0.9\textwidth]{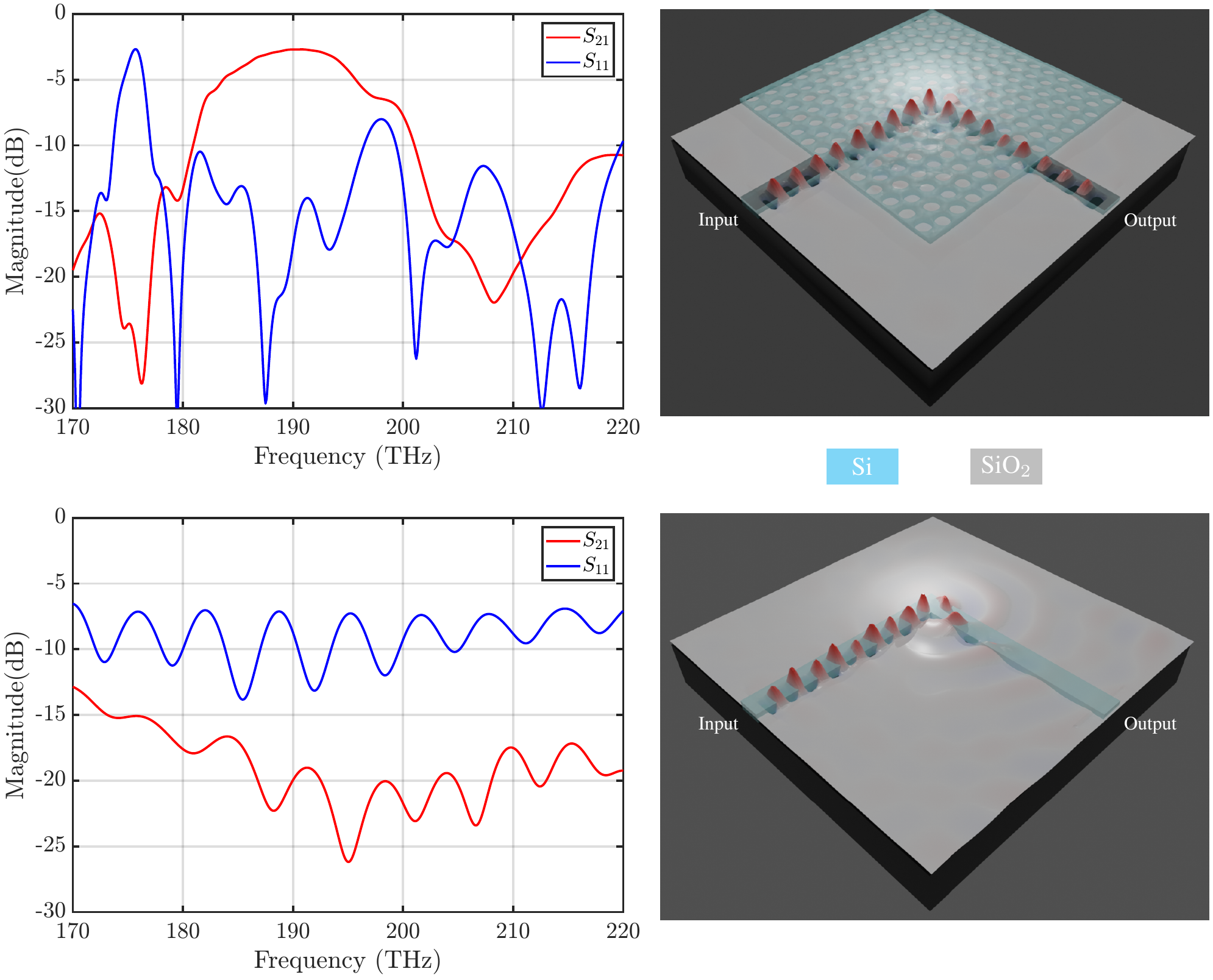}
	\caption{Simulated S parameters and electric field distributions for a 90° silicon-on-insulator waveguide designed using the morphogenetic approach (top row) and a conventional design (bottom row). The morphogenetic sample was generated with $f = 0.036$, $k = 0.065$, $d_U = 0.5$, $d_V = d_U/2$, and $\rho = 0.35$, resulting in a distribution of circular patterns with a controlled radius of $192$\,nm that gives rise to a photonic bandgap within the optical telecommunication frequency band $170–220$\,THz. The circular patterns were defined as air inclusions embedded in a silicon matrix ($\epsilon_r$ = 11.9, $\tan(\delta) = 2.5\times1e^{-4}$). The entire structure consists of a distribution of air holes ($7000 \times 7000 \times 250$\,nm$^3$) patterned on top of a SiO$_2$ substrate ($9000 \times 9000 \times 1250$\,nm$^3$), and is interfaced at the input and output regions with two ridge waveguides ($2000 \times 600 \times 250$\,nm$^3$). The conventional case consists of a 90° bent ridge waveguide with the same dimensions. The morphogenetic structure exhibits a strong field confinement through the bend, illustrating its adaptability to complex geometries without optimization.}
	\label{fig:Guide_photonic}
\end{figure}

To facilitate an accessible and reproducible proof-of-concept, we conducted our experimental validation in the 18–26\,GHz range (K-band) using simplified fabrication techniques based on laser-machined \textit{Rohacell} substrates and off-the-shelf alumina rods. This choice prioritized practicality and reproducibility over absolute performance, enabling the demonstration to be replicated with minimal infrastructure. Although this approach naturally introduces some fabrication-induced variability, these effects are common to both the morphogenetic and periodic designs, ensuring that the comparison remains meaningful. The results consistently show that the self-organized structures outperform their periodic counterparts under identical constraints.


While the present demonstration was carried out at microwave frequencies, the underlying principles of morphogenetic design are not frequency-limited. In fact, the advantages of the method become increasingly compelling at higher frequencies, where ohmic losses in metallic waveguides and fabrication inaccuracies in periodic media become dominant limitations. All-dielectric, self-organized waveguides offer a promising alternative in such regimes. To illustrate this, we extended the approach to the design of a silicon-on-insulator waveguide featuring a 90° bend and operating within the optical telecommunication frequency band. The morphogenetic sample was generated with $f = 0.036$, $k = 0.065$, $d_U = 0.5$, $d_V = d_U/2$, and $\rho = 0.35$, resulting in a distribution of circular patterns with a controlled radius of $192$\,nm that gives rise to a photonic bandgap within the desired frequency range $170–220$\,THz. As shown in Fig.~\ref{fig:Guide_photonic}, the morphogenetically generated structure, composed of air holes embedded in silicon and supported by an SiO$_2$ substrate ($\epsilon_r$ = 2.1, $\tan(\delta) = 1e^{-6}$), exhibits a broad photonic band gap and maintains strong field confinement through sharp bends.

The simulated transmission parameters clearly highlight the morphogenetic waveguide’s ability to sustain efficient propagation through nontrivial guiding geometries. Unlike conventional silicon-on-insulator designs, which suffer from substantial losses in bent geometries, the self-organized structure preserves efficient propagation along the curved path without requiring any tuning or optimization. 
	
More broadly, this work reflects a fundamental shift in electromagnetic design: from centralized, optimization-driven strategies toward decentralized, emergent approaches inspired by natural systems. By leveraging simple local rules, the morphogenetic process enables the autonomous generation of functional structures that would otherwise require case-by-case engineering effort. This perspective suggests a new paradigm that emphasizes adaptability, minimalism, and structural self-organization. To our knowledge, although the method draws direct inspiration from biological morphogenesis, no evidence has yet experimentally confirmed that reaction–diffusion mechanisms underlie the formation of natural photonic structures with band-gap properties~\cite{vukusic2009physical, burg2018self}. However, given the efficiency and simplicity of the approach demonstrated here, it remains a plausible hypothesis that similar decentralized processes could operate in biological systems where material structuring is functionally driven.

\section{Methods}

\subsection*{Design and extraction of freeform morphogenetic waveguides}

We use a bottom-up strategy to design freeform waveguides embedded in disordered media generated by reaction–diffusion dynamics. The substrate is synthesized using the Gray–Scott model~\cite{gray1983autocatalytic}, as detailed in Eq.~(1), and integrated via an explicit Euler scheme :

\begin{equation}
	\begin{aligned}
		\mathbf{U}_{n+1} &= \mathbf{U}_n + \Delta t \left( d_U \nabla^2 \mathbf{U}_n - \mathbf{U}_n \mathbf{V}_n^2 + f(1 - \mathbf{U}_n) \right),\\
		\mathbf{V}_{n+1} &= \mathbf{V}_n + \Delta t \left( d_V \nabla^2 \mathbf{V}_n + \mathbf{U}_n \mathbf{V}_n^2 - (f + k)\mathbf{V}_n \right),
	\end{aligned}
	\label{eq:gray-scott_euler}
\end{equation}

\noindent with $\Delta t = 1$, $d_U = 1$, $d_V = 0.5$, $f = 0.036$, and $k = 0.065$. We discretize the Laplacian using an isotropic 3×3 stencil, which ensures uniform diffusion in all directions. The simulation is halted before long-range order emerges, yielding robust circular spot patterns with suppressed density fluctuations and preserved disorder (see Supplementary Section 1).

The normalized concentration field $\mathbf{V}$ is binarized to define a two-phase medium, where localized peaks correspond to high-index dielectric cylinders. A direct extrusion of these regions enables the formation of cylindrical scatterers (see Supplementary Section 2).

To ensure a fair comparison between the morphogenetic medium and the triangular crystal, both structures were dimensioned using eigenmode calculations within the plane-wave expansion method. The objective was to form a band gap in the K frequency range with dielectric cylinders of radius $0.85\,\mathrm{mm}$ and a relative permittivity of $\epsilon_r = 9.8$, corresponding to commercially available alumina rods. In the morphogenetic case, an appropriate thresholding operation controls the cylinder diameter, while the spatial frequency of the generated pattern, determined by $d_U$, sets the mean inter-cylinder distance. For the triangular crystal, a lattice parameter of about $6.5$,mm places the band gap in the same frequency range. Illustrations confirming the appearance of comparable band gaps in both cases are provided in Supplementary Section 3.

To enable waveguiding, a low-density channel is created along a prescribed trajectory wide enough to support the TE\textsubscript{10} mode. In the morphogenetic case, the waveguide path is embedded during pattern formation by enforcing zero concentrations within a predefined mask. In the triangular lattice, the waveguide is carved a posteriori by removing rods intersecting the desired path. Details of both strategies are given in Supplementary Section 3.

\subsection*{Electromagnetic simulation framework}

Numerical simulations are carried out using the finite-difference time-domain method from the commercial solver \textit{CST Studio Suite}. The structures are placed in a domain bounded by perfect electric conductors along the vertical axis and absorbing perfectly matched layers (PML) in the transverse directions. A transverse electric (TE\textsubscript{10}) mode is injected via a WR-42 rectangular waveguide port aligned with the structure input. The rods are modeled as dielectric cylinders with a height of $5.33$\,mm and a radius of $0.85$\,mm. The simulation mesh is refined to ensure convergence of the $S$ parameters across the $18–26$\,GHz band. Transmission and reflection coefficients are computed for each design configuration.

\subsection*{Fabrication and experimental characterization}

The waveguides are fabricated using alumina rods with $\epsilon_r = 9.8$ and $\tan(\delta) = 2 \times 10^{-3}$, cut to a height of $5.33\,\mathrm{mm}$ to match the WR-42 waveguide cross-section. The rods are inserted into \textit{Rohacell} foam matrices ($\epsilon_r \approx 1$), laser-drilled for accurate placement. Each sample is enclosed between two aluminum plates to impose metallic boundary conditions along the vertical axis, ensuring quasi-2D propagation. TE-polarized waves are launched and collected via WR-42 ports connected to a vector network analyzer (VNA), ensuring excitation of the TE\textsubscript{10} mode. Full fabrication and measurement protocols are detailed in Supplementary Sections 4 and 5.

\subsection*{Reverse engineering and corrected designs}

Initial measurements revealed deviations from simulated responses, including a narrowing of the transmission band by approximately $1\,\mathrm{GHz}$ and a reduction in peak transmission of about $3\,\mathrm{dB}$. These discrepancies were attributed to fabrication tolerances, such as finite rod placement accuracy (within $\pm 0.2\,\mathrm{mm}$), sub-millimetric air gaps at the rod–foam interfaces, and a slightly underestimated permittivity of the \textit{Rohacell} support, possibly affected by storage conditions or ambient humidity. To assess and compensate for these effects, a reverse engineering step was carried out. In these corrected models, the permittivity of the surrounding foam was adjusted to $\epsilon_r = 1.07$ to better match the observed response (see Supplementary Section 5). The updated simulations closely reproduce the measured spectra, including the observed frequency shifts and attenuation effects. These systematic deviations were consistently observed across both morphogenetic and triangular configurations. All experimental results shown in Fig.~\ref{fig:ParamS21_90_120_S_mesu} correspond to these corrected geometries.

\section{Acknowledgments}

\subsection*{Funding}
This work is supported by the French National Research Agency (ANR JCJC MetaMorph ANR-21-CE42-0005 and LABEX $\Sigma$-LIM ANR-10-LABX-0074-01) and CNRS through the MITI interdisciplinary programs via its Exploratory Research program.\\

\subsection*{Author contributions}
Conceptualization: FC, TF\\
Methodology: FC, TF\\
Investigation: FC, TF, DRS\\
Supervision: TF, DRS, CD\\
Writing—original draft: FC, TF\\
Writing—review \& editing: FC, TF, DRS, CD\\

\subsection*{Competing interests}
The authors declare that they have no competing interests.

\subsection*{Data and materials availability}
All data needed to evaluate the conclusions in the paper are present in the paper and/or the Supplementary Materials.

\bibliography{refs}
	
\end{document}